# Engineering Polarization Rotation in Ferroelectric Bismuth Titanate

Amritendu Roy[1], Rajendra Prasad[2], Sushil Auluck[3], and Ashish Garg[1*]

[1]Department of Materials Science & Engineering, Indian Institute of Technology, Kanpur- 208016, India

[2]Department of Physics, Indian Institute of Technology, Kanpur - 208016, India

[3]National Physical Laboratory, Dr. K. S. Krishnan Marg, New Delhi-110012, India

Here, we report a combined experimental-theoretical study showing that collective application of rare earth doping on A-site and epitaxial strain to ferroelectric bismuth titanate does not lead to a very large $c$-axis polarization as reported previously. Further first principles calculations based on the examination of polarization tensor suggest that simultaneous Bi and Ti site doping could result in moderate polarization along $c$-axis of bismuth titanate which is typically a preferential axis of film growth and thus enabling $c$-axis oriented films to have appreciable polarization. This approach could also be applicable to other ferroic oxides where one can correlate the doping, epitaxial strain, and polarization to design materials compositions resulting in epitaxial films grown along desired directions yielding substantial polarization.

As a distinctive subgroup of non-centrosymmetric polar crystals, ferroelectrics exhibit switchable spontaneous polarization ($P_s$) which makes them suitable for non-volatile memory (FeRAM) applications that exploit switchable ferroelectric states to store information and retain them during power interruption.[1] Thus, $P_s$ is a crucial parameter for material selection in ferroelectric memory applications.[2] Further, the direction of $P_s$ should be such that one gets appreciable polarization in the out-of-plane direction, switching direction for a capacitor based memory device.[2] A typical approach, to maximize the polarization along the switching direction is to grow epitaxial ferroelectric thin films along the polar axis on appropriately chosen substrates or by tailoring the direction of $P_s$ by means of doping and substrate induced strain.[3-5] In this context, a curious case is of ferroelectric bismuth titanate ($Bi_4Ti_3O_{12}$ or BiT) which has long been studied for ferroelectric memory applications.

BiT has a highly anisotropic structure exhibiting strong directional dependence of its physical properties [6-8] with $P_s$ to be maximum along the *a*-axis (~55 $\mu C/cm^2$) and zero along *c*-axis.[8] However, epitaxial films of BiT have a strong propensity to grow with a preferred (001)-orientation (*c*-axis) [9-13] which expectedly exhibits negligible polarization. Thus, an alteration in the direction of $P_s$ by judicious selection of external stimuli leading to significant polarization along *c*-axis would make easily grown *c*-axis oriented BiT films, suitable for FeRAM devices. This has led to numerous studies on a variety of substrates demonstrating a combination of doping and substrate induced strain effects on polarization in BiT. [9-13] In contrast to most reports, in a startling study, Chon *et al.* [14] demonstrated a large *c*-axis polarization ($2P_r \sim 100$ $\mu C/cm^2$) in strongly *c*-axis oriented

Roy et al.Page 3 of 14

Nd-doped BiT thin films. While the authors attributed this to preferential displacement of Ti2 ion along *c*-axis upon Nd doping, one would intuitively expect a change in the crystal symmetry to justify such a large rotation in $P_s$ vector as observed in several systems having a morphotropic phase boundary.[15,16] The results of Chon et al.[14] could have been useful because, if true, these could lead to rather easily grown c-axis oriented bismuth titanate thin films on simpler substrate as compared to the substrates used for non-c-axis oriented thin films. However, subsequent study by Garg *et al.*[9] made on epitaxial films of *c*-axis and non-*c*-axis oriented Nd-doped BiT thin films on single crystalline substrates ruled out any such effect. Further studies on bulk and epitaxial films of RE doped BiT (Sm [13,17] and Pr[18,19]) continued to present conflicting reports on polarization behavior raising questions on exact role of epitaxial strain and lanthanide doping on polarization response of bismuth titanate. These led to the question whether high *c*-axis polarization achieved by Chon *et al.*[14] was indeed due to lanthanide doping and/or substrate induced strain.

      We have attempted to investigate this problem by carrying out a first-principles study on rare earth (RE: La and Nd) doped BiT to readdress the issue of $P_s$ and its direction, specifically to investigate "*what exactly happens upon RE doping and epitaxial strain in BiT*". The objectives of our present study have been two-fold: (i) to develop a theoretical approach with broader perspective in terms of relating the evolution of crystal and electronic structures and $P_s$ with doping as well as epitaxial strain and (ii) to readdress the issue of polarization rotation and its magnitude upon RE doping in BiT. In this letter, we show a small *c*-axis polarization in RE doped BiT, consistent with most experimental studies. We further, propose that simultaneous A- (Bi) and B- (Ti) site

doping by appropriate ions can lead to much larger rotation of the spontaneous polarization vector resulting significant polarization along crystallographic *c*-axis.

The compositions chosen for the present study are $Bi_4Ti_3O_{12}$ (BiT), $Bi_{3.25}La_{0.75}Ti_3O_{12}$ (BLT) and $Bi_{3.25}Nd_{0.75}Ti_3O_{12}$ (BNdT). Structural refinement of the XRD spectra (see Fig.S1 in the supplementary information) of as grown powder samples, using orthorhombic *B2cb* symmetry, yielded room temperature lattice parameters as *a* = 5.4461 Å, *b* = 5.4091 Å and *c* = 32.822 Å for BiT and *a* = 5.4317 Å, *b* = 5.4085 Å and *c* = 32.8401 Å for BLT. For BNdT, we used neutron diffraction data of Achary *et al* [20]: *a* = 5.4117 Å, *b* = 5.4003 Å and *c* = 32.834 Å. Subsequent first-principles calculations,[21] by relaxing the ionic positions such that the residual forces are small enough to be ignored, yielded relaxed ionic positions. Calculated structural parameters suggest subtle changes in the crystallographic environment in terms of polyhedral distortion and atomic displacements. Fig. 1(a) shows schematically that RE doping in BiT results in the rotation of $TiO_6$ octahedra: more significant for Ti1-O octahedra (Ti ions at the middle layer in the perovskite layer) with Ti2-O octahedra (Ti ions at the top and bottom layers in the perovskite layer) remaining almost stationary. Further, the displacement of Ti1 ion from the center of the Ti1-O octahedra along *a*-axis reduces to ~0.1 Å and ~ 0.20 Å in BLT and BNdT respectively compared to ~ 0.24 Å in case of undoped BiT. In contrast, the displacement of Ti2 ion in the top and bottom of the perovskite layer is along *c*-axis with fairly similar magnitude: ~ 0.35 Å for BiT, ~ 0.39 Å for BLT and ~ 0.36 Å for BNdT. Although the displacement of Ti2 ion in BNdT is in accordance with the value reported by Chon *et al.*[14], their argument that this displacement was the major contributor to the large *c*-axis polarization of their samples is not supported by the fact that two Ti2 ions situated in the top and bottom row of the perovskite layer are displaced in the opposite directions along *c*-axis, ruling out the possibility of contributing to *c*-axis $P_s$ in BNdT. However, as shown in subsequent sections, a calculation of $P_s$ of BiT, BLT and BNdT would further clarify the situation.

Therefore, using the Berry phase method,[22] we calculated $P_s$ with the Born effective charges ($Z(u)$) of the constituent ions in the optimized ferroelectric phase and the displacements ($u$) of the ions in the ferroelectric phase with respect to the paraelectric phase and listed them in Table 1. The advantage of calculating $P_s$ using $Z(u)$ values is that, one can get a quantitative estimate of contributions from individual ions and hence, the origin of ferroelectric response. We note that for BiT, calculated $P_s$ lies along [100]-direction and is in excellent agreement with previous experiments.[8] However, the direction of $P_s$ in RE doped BiT makes an angle of 5.37° and 5.20° with [100]-direction for La and Nd doping, respectively. This results in a small *c*-axis polarization but nowhere close to the value reported by Chon *et al.*[14] We also observe that upon RE doping, the magnitude of $P_s$ decreases with respect to that of undoped BiT, consistent with previous experimental observations on single crystals [23,24] and thin film samples.[25]

To identify the origin of such polarization reduction upon RE doping, we plotted the displacement vectors along *a*-axis for BiT, BLT and BNdT as shown in Fig. 1(b) which shows that in comparison to BiT, the total displacement of individual ions, with respect to their paraelectric positions is smaller in BLT and BNdT, particularly the displacement of O ions which contribute most to the $P_s$ of BiT. Since the displacement of Bi ions are almost similar, we can also conclude that the distortion of $TiO_6$ octahedra is reduced upon RE doping which in turn reduces the $P_s$ in the doped compositions. Similar calculations on a prototype ferroelectric, BaTiO$_3$ as a function of dopant concentration suggest a linear increase in the magnitude with no rotation of the $P_s$ (see Table S1 in the supplementary information). Our discussion, thus, clearly shows that while RE doping in BiT leads to polarization rotation, the rotation is rather small resulting in a small *c*-axis polarization.

In order to understand effect of doping on the electronic structure vis-à-vis modification in the chemical environment, we further calculated electronic structures of the pristine and the doped compositions. Fig. 2(a)-(c) show the band structure of BiT, BLT and BNDT along high symmetry directions. Our calculations suggest that pure BiT is a typical indirect band gap ($E_g \sim$ 2.29 eV (M-Z)) semiconductor as predicted by a number of previous first-principles calculations.[26-28] on orthorhombic and monoclinic structures of BiT. It is observed that the band structure is noticeably modified upon RE doping (Fig. 2(b) and (c)) with the indirect band gap, $E_g \sim$ 2.08 eV and 1.90 eV for La and Nd doping, respectively. Such modification of electronic structure upon trivalent substitution at Bi site could be explained in terms of hybridization between rare-earth metal d-state and O 2p state. Further, Bi-O hybridization which could be affected by doping can also contribute to the band gap reduction. Our calculations of cation-oxygen bond length show that some of the Bi-O bonds are shortened upon RE doping with a maximum of ~ 1.66 % and ~ 0.83 % reduction in the bond length in case of Nd and La doping, respectively. This reduction in the bond length also indicates towards a stronger hybridization between atomic orbitals which can induce wider band dispersion leading to a smaller band gap. Similar report of band gap reduction has also been made in case of La-doped BiFeO$_3$[29]. However, orbital hybridization could be better understood from the density of states plots as shown in Fig. 2(d)-(f). It is observed that the uppermost section of the valence band has mainly O 2p character. It is worth mentioning that the contribution from Ti2 ion is higher in comparison to Ti1 ion indicating stronger hybridization between Ti2 3d and O 2p orbitals. The conduction band, on the other hand, is primarily made up of Ti 3d states with significant contributions from Bi 6p and O 2p states indicating orbital hybridization. Since the contribution of O 2p bands in the conduction band is rather small, especially near the conduction band minimum, we can predict that the material has significant ionic character as well.[30] Upon doping with La, Ti2 3d band moves toward Fermi level, resulting in a reduction of the band gap. However, when we dope with Nd, Nd 5d-O 2p hybridized state appears near ~ 2.0 eV as marked by the arrow in Fig. 2(f). Such hybridization between Nd 5d and O 2p leads to significant band gap reduction in BNdT.

Though our calculations on the bulk structure predicted absence of sizable polarization rotation upon RE doping, to mimic the thin film geometry in a complete manner, we considered substrate induced strain and included that in our calculation. We applied compressive strain of 1 %, tensile strain of 1 % and 2 % on the *a-b* plane of BiT, BLT and BNdT. These strains almost cover the range of epitaxial strains that pure and doped films of BiT experience when grown on textured substrates.[9,14,25] After calculating $Z(u)$s and displacements, *u,* from the strained structures, we calculated ***P*$_s$** for BiT, BLT and BNdT structures and the results are shown in Fig. 1(c). We find that for BiT, application of epitaxial strain has a linear relation with the magnitude of ***P*$_s$**, with application of tensile strain leading to larger $P_s$. However, for BLT and BNdT, the unstrained structures demonstrate the smallest polarizations while $P_s$ increases upon application of either of compressive or tensile strain attributed to strain driven structural distortion. Moreover, while epitaxial strain does not induce any appreciable *c*-axis polarization in undoped BiT, it leads to increased *c*-axis polarization for BNdT (inset Fig. 1(c)) with $\boldsymbol{P}_{s,z}$ ~ 4.5 µC/cm$^2$ for a compressive strain of 1 % primarily due to large displacement of Bi2 ions, while other contributors are Nd, Ti1, Ti2, O1 and O3 ions. Thus it is plausible that high polarization obtained in *c*-axis oriented doped BiT films[14,17,18] could be due to extraneous reasons such as carbon retention due to solution processing of their thin films.

While we show that La and Nd doping in BiT do not yield significantly large *c*-axis $\boldsymbol{P}_s$, polarization engineering by choosing appropriate dopant resulting in significant $\boldsymbol{P}_s$ along *c*-axis, would be desirable from device application perspectives. Possibility of further enhancement in the *c*-axis component of $P_s$ could be understood from the very definition of $\boldsymbol{P}_s$ where the component along *c*-axis could be defined as:

$$P_{s,z} = Z_{31}u_x + Z_{32}u_y + Z_{33}u_z \qquad (1)$$

where, $Z_{ij}$ represents elements of Born effective charge tensors and x, y, z crystallographic directions. In case of BiT, negligible off-diagonal elements of $Z(u)$ coupled with insignificant $u_y$ and $u_z$ result in zero $P_{s,z}$ while BLT and BNdT show finite $P_{s,z}$ due to local symmetry lowering caused by doping and epitaxial strain. Thus, the strategy to have significantly large $\boldsymbol{P}_{s,z}$ can be two-fold: (i) to enhance the off-diagonal terms of $Z(u)$s and (ii) to enhance $u_y$ and $u_z$; $u_z$ in particular since it is associated with large principal element of $Z(u)$. Off-diagonal terms can be induced by increasing the structural distortion and subsequent local symmetry lowering by increasing the doping level. However, our attempts to dope at Bi2 site proved energetically unfavorable, suggesting that any further improvement in $P_{s,z}$ would involve manipulation with Nd, Ti1, O1 and O3 ions only. We found that doping with greater RE concentration does not induce large off-diagonalities in the $Z(u)$ at the respective doping site. Instead, off-diagonality appears in the $Z(u)$ of neighboring Ti1 ion with small magnitude. These small off-diagonal terms along with fractional displacements do not provide significant contributions to the overall polarization. On the other hand, we observe, O1 ion which is bonded with Ti1 ion has preferential tendency to be displaced along *c*-axis. If we can stimulate this displacement, coupled with the 33 component of $Z(u)$, O1 ion could significantly contribute to $P_{s,z}$. This hints that in addition to doping at Bi site by Nd, doping at Ti1 site would be rather advantageous since Ti1 is bonded with O1 and O3 ions in $TiO_6$ octahedron; thus an attempt to tailor the contribution of Ti1 would also associate modifications of contributions from O1 and O3 ions. Therefore, we chose to dope Ti1 site with isovalent ions Ru and Zr, having slightly larger ionic sizes[31] with compositions $Bi_{3.25}Nd_{0.75}Ti_{2.75}Ru_{0.25}O_{12}$ and $Bi_{3.25}Nd_{0.75}Ti_{2.75}Zr_{0.25}O_{12}$. Calculation of $P_{s,z}$ using eq. (1), results in significantly enhanced $P_{s,z}$ values as depicted in Fig. 1(d): Zr doping yielded a $P_{s,z}$ of ~5.71 $\mu C/cm^2$ and Ru doping yielded a relatively small $P_{s,z}$ ~ 3.71 $\mu C/cm^2$. Corresponding rotation of the $P_s$ vector away from the *a*-axis are 10.44° and 6.87°, respectively for Zr and Ru doping.

To summarize, our first-principles calculations demonstrated that it is not possible to achieve large c-axis spontaneous polarization in ferroelectric bismuth titanate by combined lanthanide substitution on the A-site as well as application of epitaxial strain. Further calculations suggested that moderate c-axis polarization can be induced by simultaneous substitution with elements such as Nd and La at Bi site and Zr and Ru at Ti site. The origin of polarization rotation lies in local symmetry lowering and consequent structural distortion upon external perturbations. Similar approach to other oxides can prove useful in determining dopants for multifunctional oxides which can lead to films with maximum polarization along specific growth directions based on the choice of substrate.

Authors thank Mathew Dawber for fruitful discussions. The work was supported by Department of Science and Technology, Govt. of India through project number SR/S2/CMP-0098/2010.

**Figure Captions**

Fig. 1 Schematic of room temperature crystal structure of bismuth titanate showing alternate perovskite and fluorite layers. The arrows indicate the direction of rotation of Ti1-O octahedra upon RE doping. (b) Displacement of constituent ions along $x$-axis in the ferroelectric phase with respect to their positions in the paraelectric phase. Cartesian $x$-axis coincides with crystallographic $a$-axis in orthorhombic and tetragonal phases of bismuth titanate. (c) Spontaneous polarization plotted as a function of epitaxial strain in BiT, BLT and BNdT. Inset shows the evolution of $c$-axis polarization as a function of epitaxial strain in BLT and BNdT and (d) Comparison of $c$-axis component of spontaneous polarization in doped and undoped BiT.

Fig. 2 Band structures and site projected density of states of (a), (d) BiT, (b), (e) BLT and (c), (f) BNdT calculated using GGA.

**List of Tables**

Table 1: Spontaneous polarization vector with Cartesian components in BiT, BLT and BNdT structures.

|      | Polarization ($\mu C/cm^2$) | | | |
| --- | --- | --- | --- | --- |
|      | $P_x$ | $P_y$ | $P_z$ | $P_S$ |
| BiT  | -58.34 | -0.15 | 0.06 | 58.34 |
| BLT  | -33.02 | 2.77  | 1.34 | 33.16 |
| BNdT | -33.79 | 1.0   | 2.95 | 33.93 |

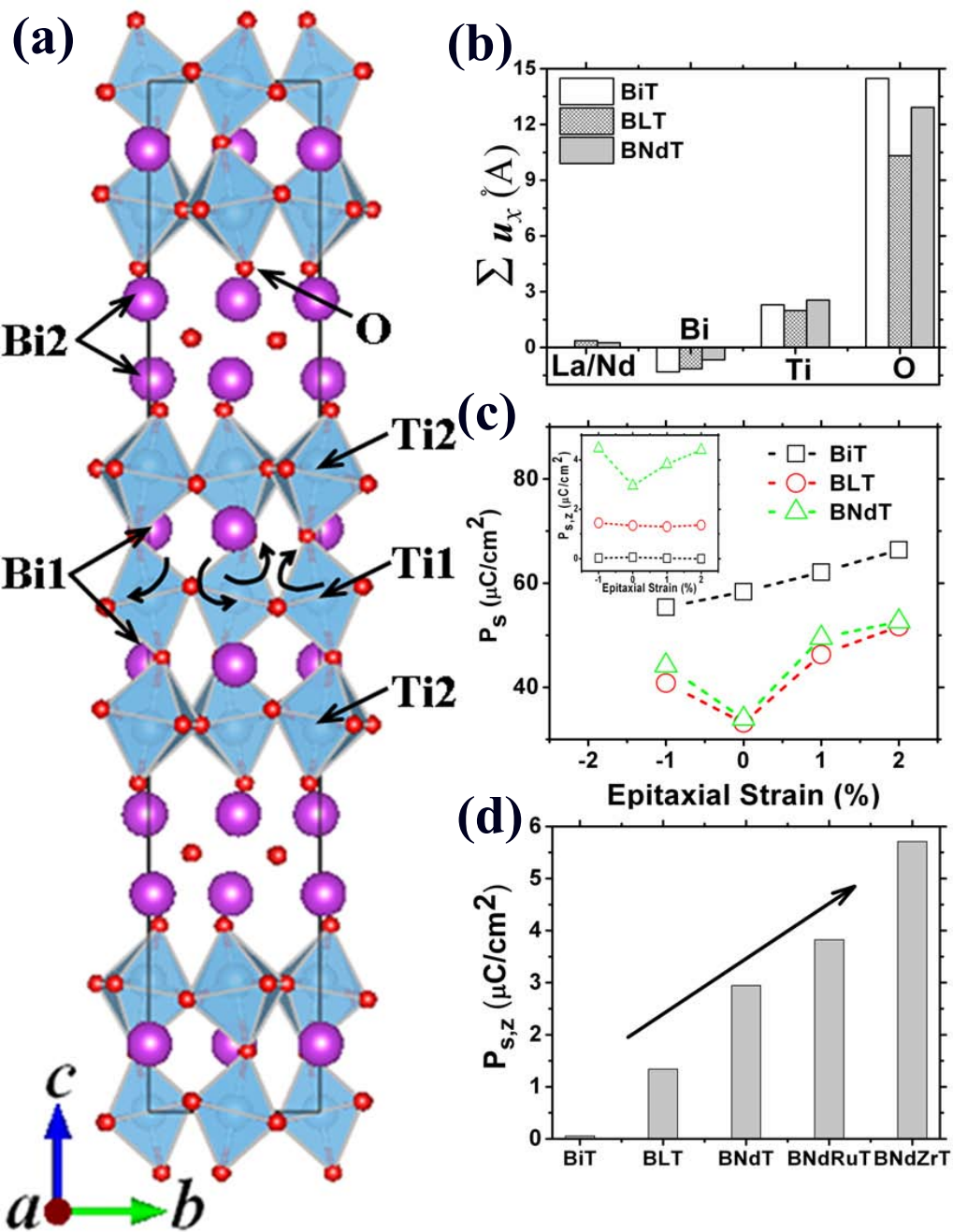

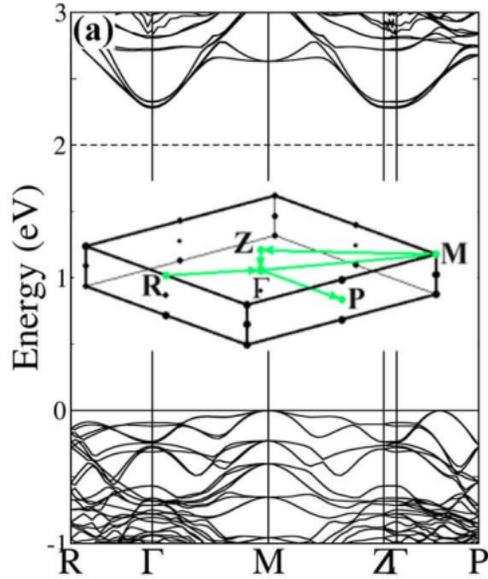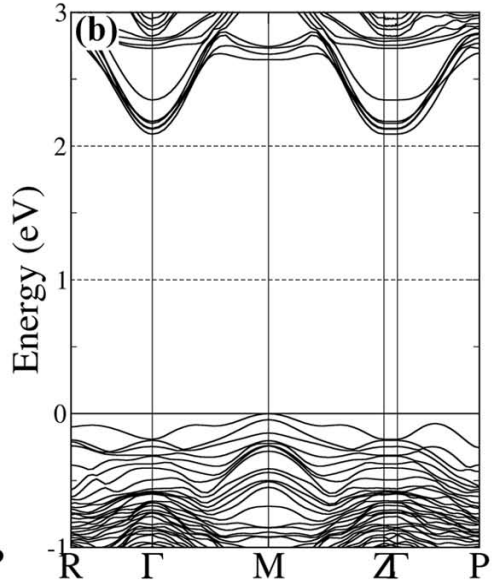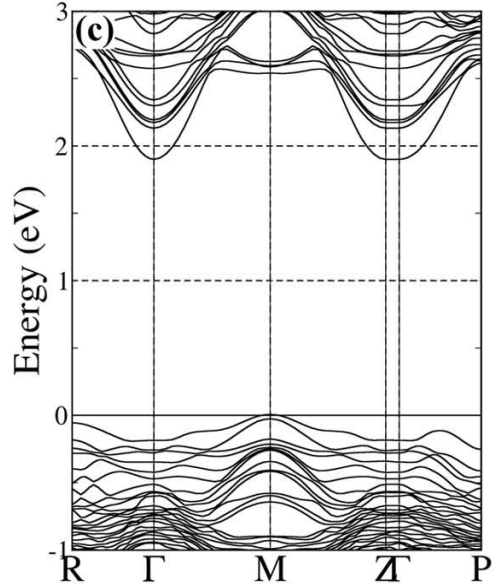
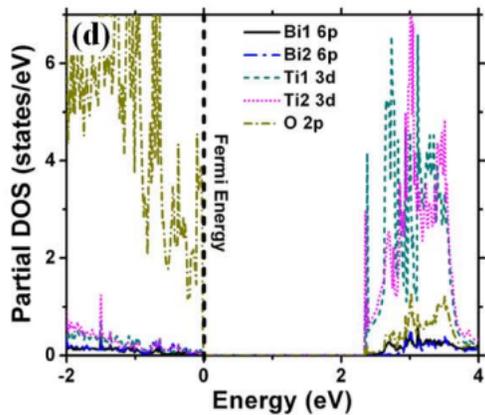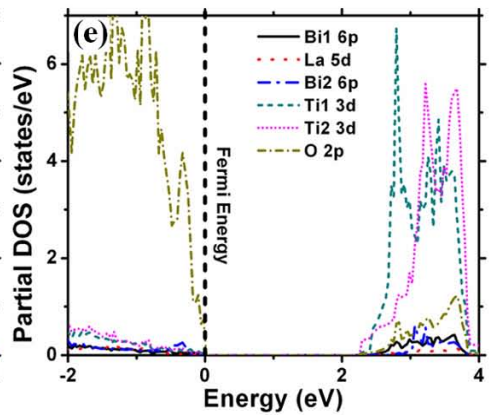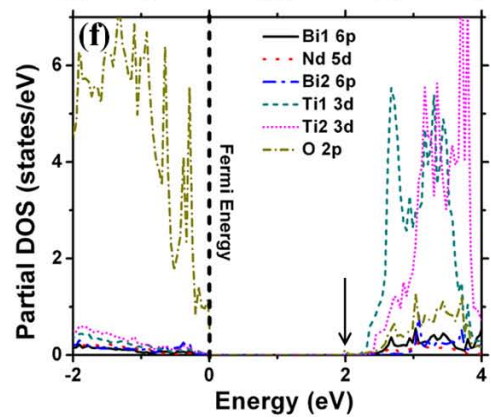